\documentclass{PoS}

\usepackage{amsmath,subfigure,latexsym,amssymb,cite}

\newcommand{\be}{\begin{equation}}
\newcommand{\ee}{\end{equation}}
\newcommand{\nn}{\nonumber}
\newcommand{\bea}{\begin{eqnarray}}
\newcommand{\eea}{\end{eqnarray}} 

\newcommand{\la}{\langle}
\newcommand{\ra}{\rangle}

\newcommand{\R}{{\kern+.25em\sf{R}\kern-.78em\sf{I} \kern+.78em\kern-.25em}}
\newcommand{\RR}{{\kern+.25em\sf{R}\kern-.6em\sf{I} \kern+.6em\kern-.25em}}
\newcommand{\N}{{\kern+.25em\sf{N}\kern-.78em\sf{I} \kern+.78em\kern-.25em}}
\newcommand{\C}{\mathbb{C}}
\newcommand{\Z}{\mathbb{Z}}

\newcommand{\ri}{{\rm i}}

\newcommand{\CP}{\C {\rm P}}
\makeatletter
\@addtoreset{equation}{section}
\makeatother

\title{Proposal for the Quantum Simulation of the $\CP(2)$ 
Model on Optical Lattices\thanks{We are indebted to D.\ Banerjee, 
L.\ Fallani, C.V.\ Kraus, M.\ Punk, E.\ Rico and S.\ Sachdev for 
helpful communication. 
This work was supported by the Schweizerischer Nationalfonds, 
the European Research Council by means of the 
European Union's Seventh Framework Programme (FP7/2007-2013)/ERC 
grant agreement 339220, the Mexican Consejo Nacional de Ciencia
y Tecnolog\'{\i}a (CONACYT) through projects CB-2010/155905 and 
CB-2013/222812, and by DGAPA-UNAM, grant IN107915.
Work in Innsbruck is partially supported by the ERC Synergy Grant UQUAM, 
SIQS, and the SFB FoQuS (FWF Project No.\ F4016-N23). 
C.L.\ was partially supported by NSERC.}}

\ShortTitle{Quantum Simulation of the $\CP(2)$ Model}
\author{Catherine Laflamme$^{\rm \ a,b}$, 
Wynne Evans$^{\rm \ c}$,
Marcello Dalmonte$^{\rm \ a,b}$, 
Urs Gerber$^{\rm \ d,e}$,
H\'{e}ctor Mej\'{\i}a-D\'{i}az$^{\rm \ d}$,
\speaker{Wolfgang Bietenholz}$^{\rm \ d}$,
Uwe-Jens Wiese$^{\rm \ c}$ and Peter Zoller$^{\rm \ a,b}$
\ \\
\vspace*{1mm}
\ \\
\ $^{\rm \ a}$ Institute for Theoretical Physics, University of Innsbruck,
A-6020, Innsbruck, Austria \vspace*{1mm} \\
\ $^{\rm \ b}$ Institute for Quantum Optics and Quantum Information \\
~~~~Austrian Academy of Sciences, A-6020 Innsbruck, Austria \vspace*{1mm} \\
\ $^{\rm c}$ Albert Einstein Center for Fundamental Physics,
Institute for Theoretical Physics \\
~~~~Universit\"{a}t Bern, Sidlerstrasse 5, CH-3012 Bern, Switzerland 
\vspace*{1mm} \\
\ $^{\rm \ d}$ Instituto de Ciencias Nucleares, Universidad Nacional 
Aut\'{o}noma de M\'{e}xico\\
~~~~A.P.\ 70-543, C.P.\ 04510 Distrito Federal, Mexico \vspace*{1mm} \\
\ $^{\rm \, e}$ Instituto de F\'{\i}sica y Matem\'{a}ticas,
Universidad Michoacana de San Nicol\'{a}s de Hidalgo\\
~~~~Edificio C-3, Apdo.\ Postal 2-82, C.P.\ 58040, 
Morelia, Michoac\'{a}n, Mexico \vspace*{3mm} \\
E-mail: \email{wolbi@nucleares.unam.mx} \\ }

\abstract{The 2d $\CP(N-1)$ models share a number of features with QCD,
like asymptotic freedom, a dynamically generated mass gap
and topological sectors. They have been formulated and analysed
successfully in the framework of the so-called D-theory, which
provides a smooth access to the continuum limit.
In that framework, we propose an experimental set-up for
the quantum simulation of the $\CP(2)$ model. It is based on
ultra-cold Alkaline-Earth Atoms (AEAs) located on the sites of an
optical lattice, where the nuclear spins represent the relevant
degrees of freedom. We present numerical results for the correlation 
length and for the real time decay of a false vacuum, 
to be compared with such a future experiment. 
The latter could also enable the exploration of $\theta$-vacua 
and of the phase diagram at finite chemical potentials,
since it does not suffer from any sign problem.}

\FullConference{The 33rd International Symposium on Lattice Field Theory\\
		14 -18 July 2015\\
		Kobe International Conference Center, Kobe, Japan}

\begin{document}

\section{Motivation}
\vspace*{-1mm}

Lattice simulations of models in quantum field theory employing
a {\em quantum system} --- {\it i.e.} analog quantum computing ---
could overcome the notorious sign problem that usually occurs if 
the Euclidean action is complex; in this approach, the phase factor
is naturally incorporated (for recent reviews, see Refs.\ 
\cite{revsignprob}).
A prominent long-term goal of this concept is the exploration of
the QCD phase diagram at finite baryon density, or a finite
vacuum angle $\theta$. In fact, within the Standard Model of 
particle physics, this is one of the major issues that remain mysterious;
so far, the sign problem has prevented reliable numerical studies.

As a step towards that goal, we present a proposal for the quantum 
simulation of the 2d $\CP (2)$ model, by means of ultra-cold 
Alkaline Earth Atoms (AEAs) trapped in an optical lattice.
Unlike previous suggestions for quantum simulations of lattice
field theory, our proposal involves an automatic extrapolation 
to the continuum limit, taking advantage of asymptotic freedom.

\vspace*{-2mm}
\section{$\CP(N-1)$ models}
\vspace*{-1mm}

The 2d $\CP(N-1)$ models \cite{Luscher} are popular toy models for QCD.
They can be considered as complex analogues of the O$(N)$ spin models, 
with a covariant derivative, as we see from the action 
in a continuous Euclidean plane,
\be  \label{act}
S [\vec z ] = \int d^{2} x \ (D_{\mu} \vec z)^{\, \dagger} \cdot
D_{\mu} \vec z - \ri \theta Q [\vec z ]\ , \quad
{\vec z} (x) \in \C^{N}, \quad |\vec z (x)| = 1, \quad %\quad N = 2,3,4 \dots
D_{\mu} = \partial_{\mu} + \frac{1}{2} ( \partial_{\mu} \vec z^{\, \dagger}
\cdot \vec z - \vec z^{\, \dagger} \cdot \partial_{\mu} \vec z ) \ ,
\ee
where $Q \in \Z$ is the topological charge, and $N = 2,3,4 \dots$. 

As a remarkable property, there is a local U(1) symmetry,
in addition to the global SU$(N)$ symmetry.
In an alternative notation, we can write the fields as $N \times N$
Hermitian projection matrices,
\be  \label{PP}
P(x) = | \vec z(x) \ra \la \vec z(x) | \ , \qquad
{\rm Tr} \, P(x) =1, \quad P(x) = P(x)^{2} = P(x)^{\dagger} \ .
\ee
The case $N=2$ corresponds to the O(3) model. For higher $N$, all 2d 
$\CP(N-1)$ models have topological sectors too (in contrast to
the higher 2d O$(N)$ models). Therefore it is natural to include a 
$\theta$-term, as it has been done in eq.\ (\ref{act}).
As further properties in common with QCD, all the 2d $\CP(N-1)$ 
models are asymptotically free, and they display a
dynamically generated mass gap. 

\vspace*{-1mm}
\subsection{D-theory formulation}

In D-theory, asymptotically free models are formulated in a space
with an additional dimension; in the weak coupling extrapolation 
towards the continuum limit, this additional direction is suppressed %removed 
by dimensional reduction \cite{Dtheory}.

In particular, the D-theory formulation of 2d $\CP(N-1)$ models
starts with 2d layers, where SU$(N)$ quantum spins are located on a
``ladder'', {\it i.e.}\ on a $L \times L'$ lattice with $L \gg L'$ 
\cite{Bern}. Hence each layer contains a set of $L'$ long quantum spin 
chains. These layers are embedded in a 3d space, which includes an 
additional $\beta$-direction.
The Hamiltonian can be written as \vspace*{-1mm}
\be  \label{Hami}
H = - J \, \sum_{\la xy \ra} \sum_{a=1}^{N^{2}-1} T_{x}^{a} T_{y}^{a \, *} \ ,
 \vspace*{-2mm}
\ee
where $T_{x}^{a} \ (T_{y}^{a \, *}) $ are spin operators in the
(anti-)fundamental representation of SU$(N)$, such that
$[ T_{x}^{a} , T_{y}^{b}] = \ri \delta_{xy} f^{abc} T_{x}^{c}$
(with the SU$(N)$ structure constants $f^{abc}$).
Here we deal with a positive coupling constant, $J>0$, which
corresponds to an anti-ferromagnetic system.

For $N=3$ and $4$, Ref.\ \cite{Troyer} pointed out that
(in the limit of zero temperature and infinite volume) this system
undergoes spontaneous symmetry breaking ${\rm SU}(N) \to {\rm U}(N-1)$,
which generates $2(N-1)$ Nambu-Goldstone bosons.
They are accommodated in the coset space of the {\em complex 
projection,} $\CP (N-1) = {\rm SU}(N)/{\rm U}(N-1)$,
such that the low energy effective description coincides 
with the $\CP (N-1)$ model.

In the notation (\ref{PP}), the D-theory continuum action
takes the form \vspace*{-1mm}
\bea
S [P] &=& \frac{1}{g^{2}} \int_{0}^{\beta} dx_{3} \int d^{2}x \ 
\mbox{Tr} \Big[ \sum_{i=1}^{2} \partial_{i} P \partial_{i} P 
+ \frac{1}{c^{2}} \partial_{3} P \partial_{3} P \Big]
- \ri \, \theta \, Q[P] \ ,  \nn \\
\ri \theta Q[P] &=& \frac{1}{\pi} \int_{0}^{\beta} dx_{3} \int d^{2} x \
\mbox{Tr} \, [ P \partial_{1} P \partial_{3} P ]  \ ,
\quad g^{2} = \frac{c}{\rho_{\rm s} \, L'} \ ,
\label{cp23}
\eea
where $c$ is the spin wave velocity, and $\rho_{\rm s}$
the spin stiffness.

If we return to a finite $L \times L'$ lattice on the spatial
layers, asymptotic freedom implies that the (spatial) correlation length 
$\xi$ (the inverse mass of the quasi Nambu-Goldstone bosons)
diverges exponentially when $L'$ becomes large, \vspace*{-4mm}
\be  \label{xiexp}
\xi \propto e^{\mbox{const.} \ L'} \ ,
\vspace*{-1mm}
\ee
where we assume $L' \ll L, \, \beta$. This divergence leads to
dimensional reduction; ironically, as $L'$ grows,
it becomes negligible, since $\xi \gg L'$. Thus we recover the
2d $\CP (2)$ or $\CP (3)$ model, with 
%the action (\ref{cp23}), and 
$\theta = L' \pi$ \cite{Bern} (since the integrand of the 
expression for $Q[P]$ is constant in $x_{2}$).

In view of the prospects to implement the SU$(N)$ quantum spin
system experimentally \cite{preprint}, to be discussed in the 
next section, it is important to explore this exponential
grow explicitly. 
\begin{figure}
\vspace*{-9cm}
\hspace*{2.3cm}
\includegraphics[angle=0,width=1.21\linewidth]{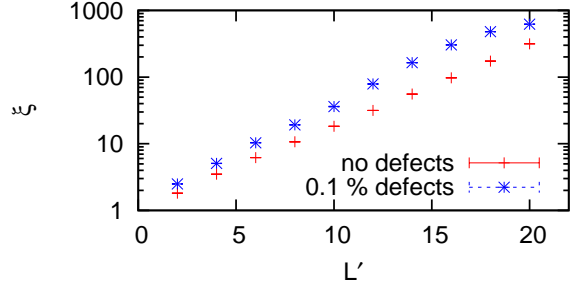}
\vspace*{-2.15cm}
\caption{The correlation length $\xi$ as a function of $L'$, for an
SU(3) quantum spin model on a $\beta \times L \times L'$ lattice, 
at $L = 1500$ and $\beta J = 1000$. The red crosses (blue asterisks) 
refer to the setting when all sites are occupied (with $0.1 \, \%$ of
random distributed empty sites). Error bars are included, but hardly
visible. The results for $\xi$ and for the second moment correlation 
length $\xi_{2}$ agree up to tiny differences, which are not visible 
either. With or without defect sites,
$\xi$ grows exponentially in $L'$,
which confirms eq.\ (\protect\ref{xiexp}) and therefore asymptotic 
freedom; we further see that a moderate $L' \approx 10$ is sufficient
for dimensional reduction.}
\label{Lprimexi}
\end{figure}
Figure \ref{Lprimexi} shows simulation results, which 
were obtained with a loop cluster algorithm \cite{loopclu} at $L=1500$ 
and $\beta J = 1000$. The boundary conditions are open in the (short)
$L'$-direction, as in the experiment, and periodic in the long
directions (where it hardly matters).
We obtain ${\rm const.}=0.270784$, and
observe that $L' \approx 10$ is sufficient for the
dimensional reduction to set in, essentially. In fact, such a
number of coupled quantum spin chains is experimentally realistic.

In an experiment it may happen that a few sites
in the optical lattice remain unoccupied. Frequent repetition of
this experiment restores translation invariance statistically,
but the correlation length is enhanced. Numerical data for this
setting, with $0.1 \%$ of such defect sites are also shown in 
Figure \ref{Lprimexi}.
Again there is clear evidence that $\xi$ grows exponentially in $L'$
(as long as finite size effects are small),
in accordance with relation (\ref{xiexp}), which reflects asymptotic 
freedom. Dimensional reduction sets is even earlier in this case.

\vspace*{-2mm}
\section{Experimental set-up}
\vspace*{-1mm}

\begin{figure}
\vspace*{-1.5mm}
\begin{center}
\includegraphics[angle=0,width=.63\linewidth]{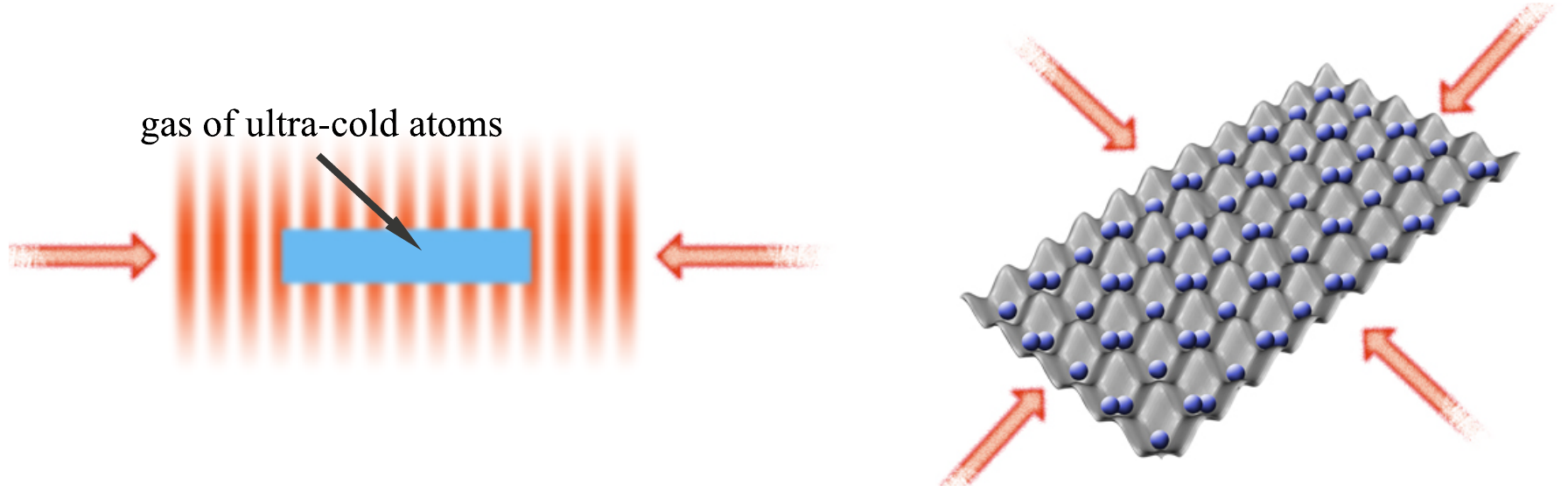}
\end{center}
\vspace*{-6mm}
\caption{Illustration of optical lattices, where the sites 
correspond to the nodes of standing laser waves.} 
\vspace*{-1mm}
\label{optlat}
\end{figure}

The goal is to implement the Hamiltonian (\ref{Hami}) by ultra-cold
fermionic AEAs on the sites of an optical lattice. 
The latter is formed by the nodes of superimposed standing laser waves, 
as sketched in Figure \ref{optlat}, with a spacing in the $\mu$m magnitude.
The temperature is of the order of nK, which keeps the atoms in
their electronic ground state. 
For fermionic AEAs, the electron and nuclear spins decouple 
almost completely in the ground state manifold, which
excludes spin changing collisions. Moreover, 
in an external magnetic field, the interactions 
between the Zeeman states are SU$(N)$ symmetric,
where $N \leq 2I+1$ and $I$ is the nuclear spin \cite{Gorsh}, 
and the total spin is conserved. 
This can be implemented up to $N=10$, {\it e.g.}\ with \ $^{87}$Sr
atoms \cite{Gorsh}.

\begin{figure}
\vspace*{-4mm}
\begin{center}
\includegraphics[angle=0,width=.53\linewidth]{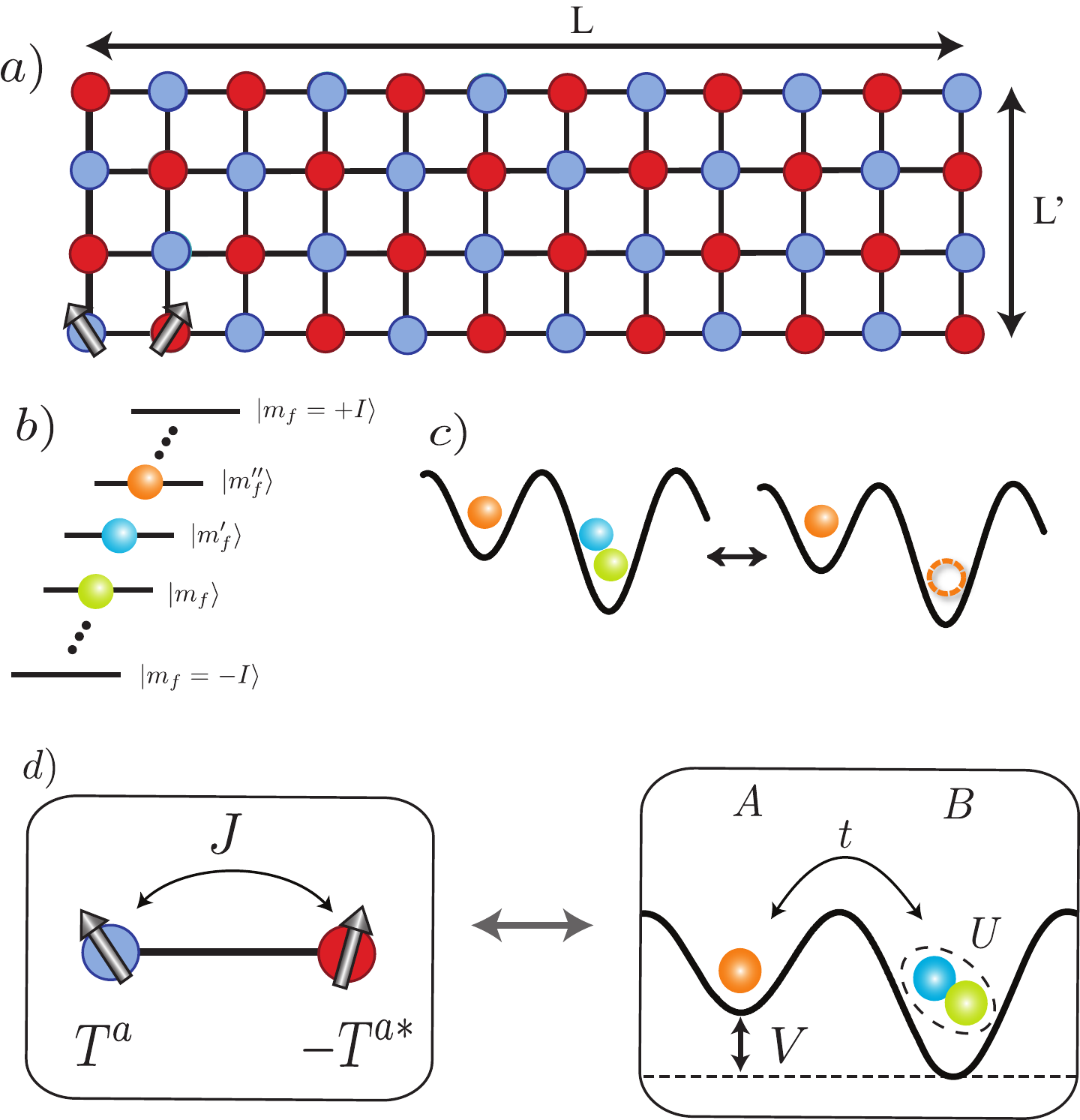}
\end{center}
\vspace*{-3mm}
\caption{An illustration of the experimental set-up:
a) SU$(N)$ spins on a bipartite $L' \times L$ lattice, $L\gg L'$.
b) The $N \leq 2I +1$ hyperfine states of an AEA, in an external
magnetic field which induces Zeeman splitting.
c) On sub-lattice $A$ the SU$(N)$ quantum spins are in the
fundamental representation, and we identify $c_{x} = d^{\dagger}_{x}$.
On sub-lattice $B$ they are in the anti-fundamental representation,
and $c_{x} = d_{x}$. Hence the exchange of the staggered sub-lattices
$ A \leftrightarrow B$ corresponds to a particle-hole transformation.
d) The nearest-neighbour interactions 
between $T^{a}_{x \in A}$ and $-T^{b}_{y \in B}$.}
\label{setup}
\vspace*{-3mm}
\end{figure}

Next we re-write the spin operators in terms of fermionic bilinears, 
composed of $d_{x}$ and $d^{\dagger}_{x}$,
\be
T_{x}^{a} = d^{\dagger}_{xm} \lambda_{mm'}^{a} d_{xm'} \ , \quad
- T_{x}^{a~*} = -d^{\dagger}_{xm} \lambda_{mm'}^{a~*} d_{xm'} \ ,
\ee
where $m,\, m'= 1 \dots N$ label the states, and $\lambda^{a}$ are 
generalised Gell-Mann matrices, ${\rm Tr} \, [\lambda^{a}, \lambda^{b}]
= \delta^{ab}$. Now the Hamiltonian is split into a hopping term
and a potential,
\be
H = H_{t} + H_{U} \ , \quad
H_{t} = -t \sum_{\la xy \ra,m} \Big( c^{\dagger}_{xm} c_{ym} +
c^{\dagger}_{ym} c_{xm} \Big) \ , \quad
H_{U} = \frac{U}{2} \sum_{x} n_{x}(n_{x}-1) + V \sum_{x \in A} n_{x} \ , \quad
\ee
where the operator $c_{xm}$ annihilates the nuclear spin level
$m \in \{ -I \dots I \}$ at site $x$, $t$ is the hopping parameter,
$U$ the on-site interaction, $n_{x} = \sum_{m} c^{\dagger}_{xm} c_{xm}$
the occupation number, and $V$ is the energy offset between two
staggered sub-lattices, which we denote as $A$ and $B$.
This system is illustrated in Figure \ref{setup}, and the caption
explains the relation between the operators $c_{xm}$ and $d_{xm}$.
The initial state should be prepared with one AEA on each
site of sub-lattice $A$, which corresponds to one fermion;
on each site of sub-lattice $B$ there are $N-1$ AEAs,
corresponding to $N-1$ fermions, or one hole.

The hopping parameter $t$ fixes the coupling constant $J$,
to be tuned by varying the energy offset $V$.
At strong coupling, $t \ll U, \, V$, the system is essentially
in the eigenstates of $H_{U}$, with virtual tunnelling due to
the SU$(N)$ exchange terms in $H_{t}$. A hopping parameter expansion
up to O$((t/U)^{2})$ reproduces the form of $H$ in eq.\ (\ref{Hami})
\cite{Auer}, with \vspace*{-2mm}
\be
J = \frac{t^{2}U}{[-V + U(N-3)]\, [V - U(N-1)]} \ .
\ee

The preparation of the initial state proceeds as follows:
first one fills each site with $N$ AEAs, where each level $m$
is occupied once\footnote{A non-uniform occupation of the Zeeman 
states may also be intended, since it captures the $\CP (N-1)$ 
model at finite density.} 
(this can be achieved by optical pumping \cite{Scazza}).
Then each site is split adiabatically into a 
double-well, forming the sub-lattices $A$ and $B$,
cf.\ Figure \ref{setup}. The barrier
is tuned to match the quantum dynamics according to
the Hamiltonian $H$; this has already been realized for bosonic
alkaline atoms \cite{Nasc}. Based on experimental experience,
we expect this to be feasible up to large $L = {\rm O}(1000)$ and
$L' \approx 12$; according to our results in Figure \ref{Lprimexi},
this is sufficient for dimensional reduction to set in.
The results for the correlation length $\xi$ can be confronted 
with experimental measurements by means of Bragg spectroscopy 
or noise correlation \cite{Bloch}. 

\vspace*{-2mm}
\section{Phase transition at $\theta = \pi$ and false vacuum decay}
\vspace*{-1mm}

An odd number $L'$ implies $\theta = \pi$, where a first
order phase transition and the spontaneous breaking of the
C (charge conjugation) symmetry is expected \cite{Seiberg}; 
Refs.\ \cite{Bern} provide numerical evidence for this scenario.

\begin{figure}
\vspace*{-3mm}
\begin{center}
\includegraphics[angle=0,width=.51\linewidth]{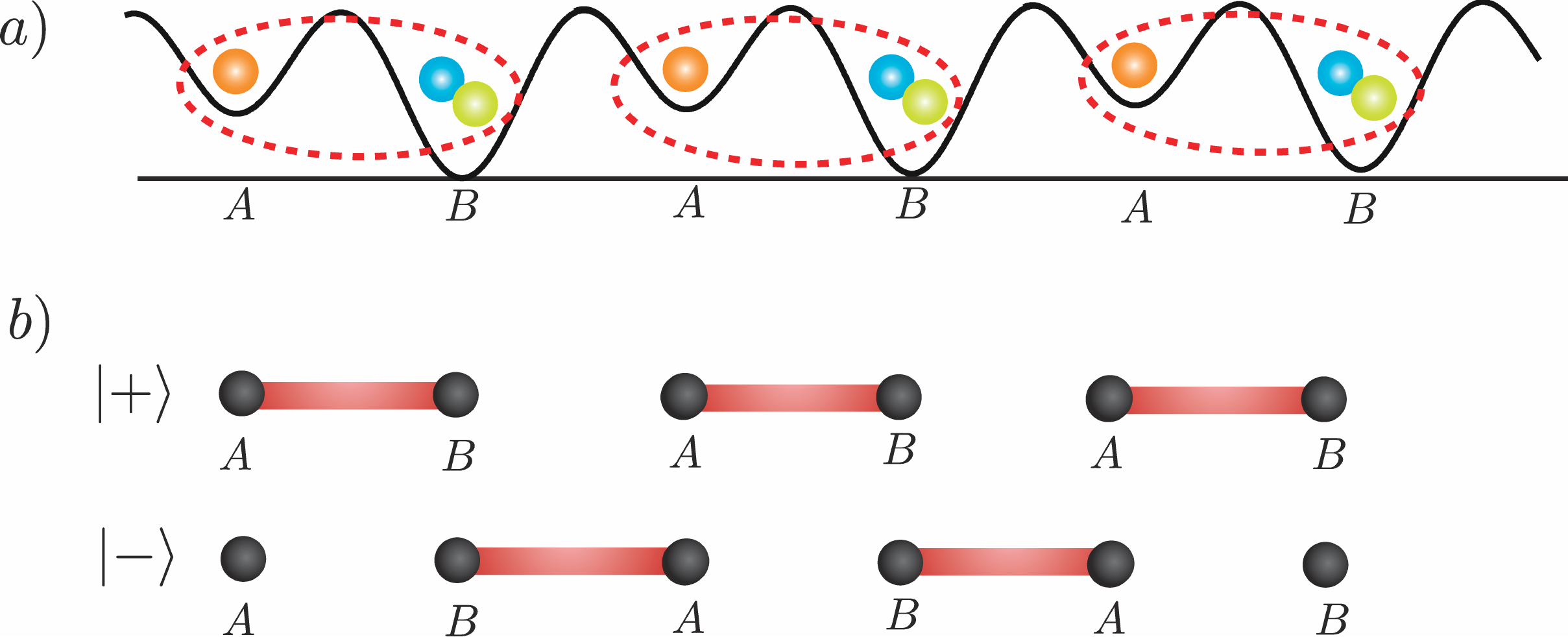}
\end{center}
\vspace*{-4mm}
\caption{An illustration of the two degenerate ground states
at $\theta =\pi$. In this case, there is dimerisation (in an ambiguous 
way), and C symmetry is broken.}
\label{bondfig} 
\vspace*{-2mm}
\end{figure}

In the experiment, a C transformation corresponds to a shift
in the $L$-direction by one lattice spacing. If $x$ ($x + \hat 1$)
belongs to sub-lattice $A$ ($B$), then this shift transforms
$T_{x}^{a} \to - T_{x + \hat 1}^{a~*}$, and $-T_{x}^{a} \to T_{x + \hat 1}^{a~*}$.
The order parameter for C symmetry \cite{RS},
\be  \label{dimer}
D = \sum_{x \in A} \la T_{x}^{a} T_{x + \hat 1}^{a~*} -
T_{x}^{a} T_{x - \hat 1}^{a~*} \ra \ ,
\ee
detects {\em dimerisation.} At $\theta =0$, C invariance holds,
and $D=0$. It breaks at $\theta = \pi$, where there are
two degenerate ground states with $\pm D$, which we denote as
$| \pm \ra$. These ground states may be distinguished by bonds
between nearest-neighbour sites, which can be set in two ways,
as shown in Figure \ref{bondfig}.
For ultra-cold atoms, the singlets that contribute to $D$
can be measured by spin changing collisions \cite{Bloch,Paredes}.

\begin{figure}
%\vspace*{-1mm}
\begin{center}
\includegraphics[angle=0,width=.8\linewidth]{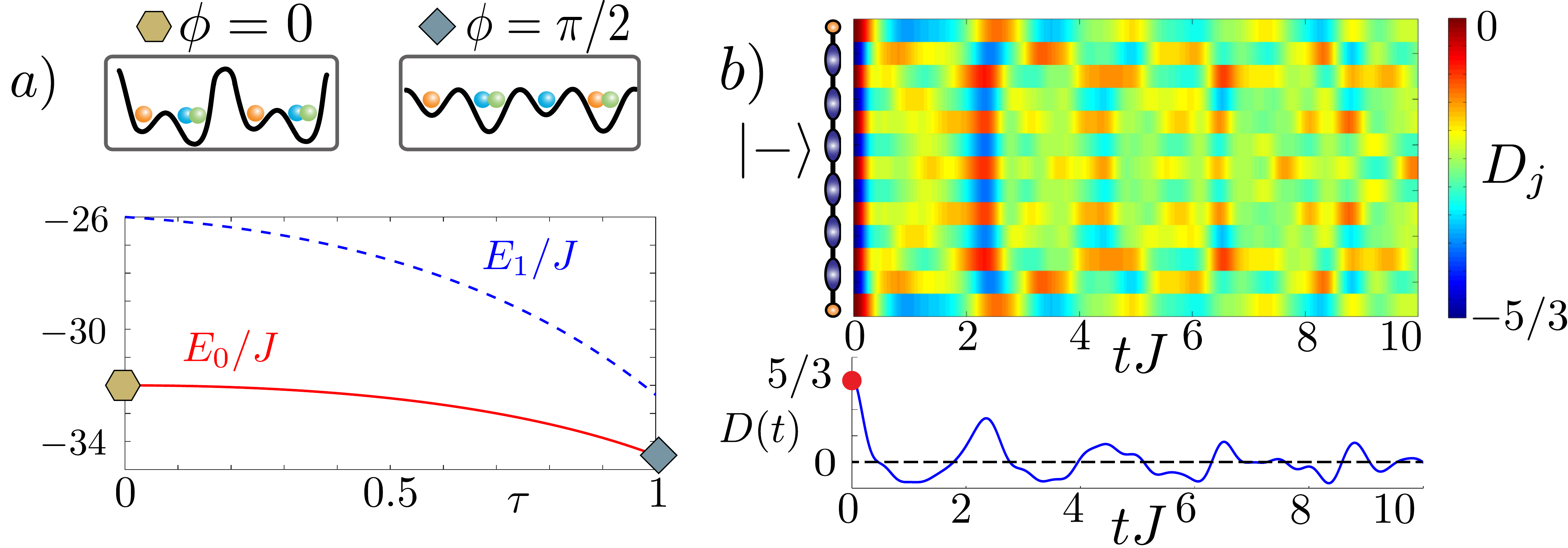}
\end{center}
\vspace*{-5mm}
\caption{The time evolution of the first two energy eigenvalues,
$E_{0}/J$ and $E_{1}/J$ (on the left), and of the 
dimerisation $D(t)$ (on the right),
if one starts from total dimerisation and switches on the hopping 
parameter $t$ adiabatically. Then the Hamiltonian (\protect\ref{hamiphi}) 
drives the dynamics, turning the initial vacuum state $|-\ra$ into
a false vacuum, which decays such that $D(t)$ decreases and oscillates, 
as shown on the right.}
\label{evolution}
\vspace*{-1.5mm}
\end{figure}

At last we consider a dynamical process for a single spin chain
of even length $L$:
starting from total dimerisation, one turns on the hopping
parameter $t$ adiabatically. We describe the gradual modification
by a parameter $\tau$, which increases from $0$ to $1$,
such that the dynamics is driven by the Hamiltonian
\be  \label{hamiphi}
H(\tau ) = \tau H - (1-\tau ) \ J \, \sum_{x \in A}
T_{x}^{a} T_{x + \hat 1}^{a~*} \ .
\ee
Numerical results for the evolution of the first two energy eigenvalues,
$E_{0}$ and $E_{1}$, are shown in Figure \ref{evolution} on the left.
Their evaluation also reveals the time dependent dimerisation $D(t)$. 
The latter is shown in the right panel of Figure \ref{evolution}, 
along with a dimerisation map in the state $|- \ra$. 
The evolution turns it into a false 
vacuum, which performs an (incomplete) decay towards the true vacuum 
$|+ \ra$, such that $D(t)$ decreases in an oscillatory manner. 
For a similar study of the real time dynamics of coupled
bosonic spin chains, we refer to Ref.\ \cite{Rosario}.

This time evolution has been computed by the exact diagonalisation
of $H (\tau )$ at $L=14$. It corresponds to the real time evolution 
of a false vacuum in the $\CP (2)$ model, which cannot be obtained
with classical Monte Carlo simulations, due to the sign problem.
The experimental set-up described here should enable a quantum
simulation, which can be compared to the results in Figure 
\ref{evolution}, and which can be extended to large $L$.

\vspace*{-2mm}
\section{Summary}
\vspace*{-1mm}

We have described a proposal for the quantum simulation of
$\CP (N-1)$ models by ultra-cold AEAs trapped in an optical 
lattice. They represent a model of SU$(N)$ quantum spins,
with $N \leq 2 I +1$, where $I$ is the nuclear spin.
This system corresponds to the 3d D-theory formulation of the 
$\CP (N-1)$ model, where dimensional reduction leads directly
to the continuum limit of the 2d model, thanks to asymptotic freedom.
Our results for the correlation length at $N=3$ show that for a realistic 
system size, dimensional reduction leads to the 2d continuum
$\CP (2)$ model.  

Experimental tools for the ground state preparation in such systems, 
and also for its adiabatic modification, do already exist 
\cite{Gorsh,Scazza,Nasc,Bloch,Paredes}. 
We discussed the dynamics of C symmetry restoration, which corresponds 
to a real time evolution in the $\CP (2)$ model. For small systems it
was evaluated by the diagonalisation of the Hamiltonian; for
larger systems it can be measured by the experiment, which acts as
an analog quantum computer.


\vspace*{-2mm}

\begin{thebibliography}{10}
\vspace*{-1mm}

\bibitem{revsignprob} U.-J.\ Wiese,
% Ultracold Quantum Gases and Lattice Systems: Quantum Simulation of Lattice Gauge Theories
Annalen Phys.\ 525 (2013) 777.\\
%arXiv:1305.1602 [quant-ph] 
E.\ Zohar, J.I.\ Cirac and B.\ Reznik,
% Quantum Simulations of Lattice Gauge Theories using 
%Ultracold Atoms in Optical Lattices
arXiv:1503.02312 [quant-ph]

\bibitem{Luscher} A.\ D'Adda, M.\ L\"{u}scher and P.\ Di Vecchia, 
Nucl.\ Phys.\ B146 (1978) 63.

\bibitem{Dtheory} S.\ Chandrasekharan and U.-J.\ Wiese,
% Quantum link models: A Discrete approach to gauge theories
Nucl.\ Phys.\ B492 (1997) 455.\\
R.\ Brower, S.\ Chandrasekharan and U.-J. Wiese,
% QCD as a quantum link model
Phys.\ Rev.\ D60 (1999) 094502.\\
R.\ Brower, S.\ Chandrasekharan, S.\ Riederer and U.-J.\ Wiese,
% D theory: Field quantization by dimensional reduction of discrete variables
Nucl.\ Phys.\ B693 (2004) 149.

\bibitem{Bern} B.B.\ Beard, M.\ Pepe, S.\ Riederer and U.-J.\ Wiese,
Phys.\ Rev.\ Lett.\ 94 (2005) 010603.\\
S.\ Riederer, %``D-Theory Formulation of 
%Quantum Field Theories and Application to CP($N-1$) Models'',
Ph.D.\ thesis, Universit\"{a}t Bern, 2006.

\bibitem{Troyer} K.\ Harada, N.\ Kawashima and M.\ Troyer,
Phys.\ Rev.\ Lett.\ 90 (2003) 117203.

\bibitem{preprint} C.\ Laflamme {\it et al.},
%, W.\ Evans, M.\ Dalmonte, U.\ Gerber, 
%H.\ Mej\'{\i}a-D\'{\i}az, W.\ Bietenholz, U.-J.\ Wiese and P. Zoller,
%CP(N-1) Quantum Field Theories with Alkaline-Earth Atoms in Optical Lattices
arXiv:1507.06788 [quant-ph].

\bibitem{loopclu} H.G.\ Evertz, G.\ Lana and M.\ Marcu,
Phys.\ Rev.\ Lett.\ 70 (1993) 875.\\
U.-J.\ Wiese and H.-P.\ Ying, Z.\ Phys.\ B93 (1994) 147.
%N.\ Kawashima and K.\ Harada, JPSJ 73 (2004) 1379.

\bibitem{Gorsh} A.V.\ Gorshkov {\it et al.}, Nature Phys.\ 6 (2010) 289.

\bibitem{Auer} A.\ Auerbach, Interacting Electrons and Quantum 
Magnetism (Springer, 1994).

\bibitem{Scazza} G.\ Pagano {\it et al.}, Nature Phys.\ 10 (2014) 198.
F.\ Scazza {\it et al.}, Nature Phys.\ 10 (2014) 779.

\bibitem{Nasc} S.\ Nascimb\`{e}ne {\it et al.}, Phys.\ Rev.\
Lett.\ 108 (2012) 205301.

\bibitem{Bloch} I.\ Bloch, J.\ Dalibard and S.\ Nascimb\`{e}ne,
Nature Phys.\ 8 (2012) 267.

\bibitem{Seiberg} N.\ Seiberg, Phys.\ Rev.\ Lett.\ 53 (1984) 637.

\bibitem{RS} N.\ Read and S.\ Sachdev, Phys.\ Rev.\ Lett.\ 62 (1989) 1694.

\bibitem{Paredes} B.\ Paredes and I. Bloch, Phys.\ Rev.\ A77 (2008) 023603.

\bibitem{Rosario} S.F.\ Caballero-Ben\'{\i}tez and R.\ Paredes,
Phys.\ Rev.\ A85 (2012) 023605.
%; Phys.\ Rev.\ A87 (2013) 053603.

\end{thebibliography}
\end{document}